\documentclass[manuscript]{aastex}
\usepackage{lineno}

\shorttitle{Searches for Point-like and extended neutrino sources close to the Galactic Centre using the ANTARES neutrino Telescope}
\shortauthors{ANTARES collaboration}

\begin{document}

\title{Searches for Point-like and extended neutrino sources close to the Galactic Centre using the ANTARES neutrino Telescope}

\author{S.~Adri\'an-Mart\'inez\altaffilmark{1},
A.~Albert\altaffilmark{2},
M.~Andr\'e\altaffilmark{3},
M.~Anghinolfi\altaffilmark{4},
G.~Anton\altaffilmark{5},
M.~Ardid\altaffilmark{1},
J.-J.~Aubert\altaffilmark{6},
B.~Baret\altaffilmark{7},
J.~Barrios-Mart\'i\altaffilmark{8},
S.~Basa\altaffilmark{9}
% ...
}

\altaffiltext{1}{\scriptsize{Institut d'Investigaci\'o per a la Gesti\'o Integrada de les Zones Costaneres (IGIC) - Universitat Polit\`ecnica de Val\`encia. C/  Paranimf 1 , 46730 Gandia, Spain.}}
\altaffiltext{2}{\scriptsize{GRPHE - Institut universitaire de technologie de Colmar, 34 rue du Grillenbreit BP 50568 - 68008 Colmar, France}}
\altaffiltext{3}{\scriptsize{Technical University of Catalonia, Laboratory of Applied Bioacoustics, Rambla Exposici\'o,08800 Vilanova i la Geltr\'u,Barcelona, Spain}}
\altaffiltext{4}{\scriptsize{INFN - Sezione di Genova, Via Dodecaneso 33, 16146 Genova, Italy}}
\altaffiltext{5}{\scriptsize{Friedrich-Alexander-Universit\"at Erlangen-N\"urnberg, Erlangen Centre for Astroparticle Physics, Erwin-Rommel-Str. 1, 91058 Erlangen, Germany}}
\altaffiltext{6}{\scriptsize{CPPM, Aix-Marseille Universit\'e, CNRS/IN2P3, Marseille, France}}
\altaffiltext{7}{\scriptsize{APC, Universit\'e Paris Diderot, CNRS/IN2P3, CEA/IRFU, Observatoire de Paris, Sorbonne Paris Cit\'e, 75205 Paris, France}}
\altaffiltext{8}{\scriptsize{IFIC - Instituto de F\'isica Corpuscular, Edificios Investigaci\'on de Paterna, CSIC - Universitat de Val\`encia, Apdo. de Correos 22085, 46071 Valencia, Spain}}
\altaffiltext{9}{\scriptsize{LAM - Laboratoire d'Astrophysique de Marseille, P\^ole de l'\'Etoile Site de Ch\^ateau-Gombert, rue Fr\'ed\'eric Joliot-Curie 38,  13388 Marseille Cedex 13, France}}
% ...

\begin{abstract}

A search for cosmic neutrino sources using six years of data collected by the ANTARES neutrino telescope has been performed. Clusters of muon neutrinos over the expected atmospheric background have been looked for. No clear signal has been found. The most signal-like accumulation of events is located at equatorial coordinates RA=$-$46.8$^{\circ}$ and Dec=$-$64.9$^{\circ}$ and corresponds to a 2.2$\sigma$ background fluctuation. In addition, upper limits on the flux normalization of an E$^{-2}$ muon neutrino energy spectrum have been set for 50 pre-selected astrophysical objects. Finally, motivated by an accumulation of 7 events relatively close to the Galactic Centre in the recently reported neutrino sample of the IceCube telescope, a search for point sources in a broad region around this accumulation has been carried out. No indication of a neutrino signal has been found in the ANTARES data and upper limits on the flux normalization of an E$^{-2}$ energy spectrum of neutrinos from point sources in that region have been set. The 90\% confidence level upper limits on the muon neutrino flux normalization vary between 3.5 and 5.1$\times$\textrm{10$^{-8}$ GeV$\,$cm$^{-2}$s$^{-1}$}, depending on the exact location of the source.

\end{abstract}


\section{Introduction}

The scientific motivation of neutrino telescopes relies on the unique properties of neutrinos, which can be used to observe and study the high-energy Universe. Cosmic rays or high-energy photons have intrinsic limitations: the mean free path of gamma-ray photons strongly depends on their energy, while magnetic fields deflect cosmic rays, diluting the information about their origin. Neutrinos are stable, neutral and weakly interacting particles, and therefore they point directly back to their origin. In addition, neutrinos are expected to originate at the same locations where the acceleration of cosmic rays and the associated production of high-energy photons take place \cite[]{RepProg, Stecker, Bednarek}. The first evidence of such a cosmic neutrino signal has recently been reported by IceCube ~\cite[]{IceCube1,IceCube2}, including in particular a cluster of events close to the Galactic Centre. The better view of the Southern Hemisphere afforded by the ANTARES neutrino telescope, due to its location in the Mediterranean Sea,  provides an increased sensitivity to galactic sources of neutrinos with energies $<$ 100 TeV. This is particularly important in order to interpret the cluster of events observed by IceCube close to the Galactic Centre.

In this paper the results of the search for point sources with the data gathered between 2007 and 2012 with the ANTARES neutrino telescope are presented. After a brief description of the apparatus, the data selection and the corresponding detector performance are presented in Sections~\ref{ANTARESdet} and~\ref{performance}, respectively. In Section~\ref{method}, the search method is explained. The results of the full-sky and candidate sources searches are presented in Section~\ref{results}. The implications on some recent interpretations of the IceCube results are discussed in Section~\ref{Implications}. Finally, the conclusions are given in Section~\ref{conclusions}. 

\section{The ANTARES neutrino telescope and data selection}\label{ANTARESdet}

ANTARES is an underwater neutrino telescope located 40 km to the South of Toulon (France) in the Mediterranean Sea (42$^\circ$ 48' N, 6$^\circ$ 10' E) \cite[]{AntDetect}. It is made of 12 slender lines spaced by about 65 m, anchored on the seabed at 2475 m depth and maintained vertical by a buoy. Each line of 350 m active length comprises 25 floors spaced regularly, each housing 3 photomultiplier tubes (PMTs) looking downward at an angle of 45$^\circ$. The detection principle is based on the observation of the Cherenkov light induced by muons produced in charged current interactions of high energy neutrinos inside or near the detector volume. Some of the emitted photons produce a signal in the PMTs (``hits'') with the corresponding charge and time information. The hits are used to reconstruct the direction of the muon. In addition, other neutrino signatures such as cascade events are also detected and reconstructed. The current analysis uses muon tracks only, which offer a better angular resolution and larger volume than cascades caused by showering events. 

High quality runs are selected from data between January 29, 2007 to December 31, 2012. This measurment period corresponds to a total livetime of 1338 days, which is an increase of 70\% compared to the previous ANTARES point-source analysis \cite[]{PrevAna}. 

Triggered events are reconstructed using the time and position information of the hits by means of a maximum likelihood (ML) method \cite[]{AartThesis}. The algorithm consists of a multi-step procedure to fit the direction of the reconstructed muon by maximising the ML-parameter $\Lambda$, which describes the quality of the reconstruction. In addition, the uncertainty of the track direction angle, $\beta$, is calculated. This calculation is estimated from the uncertainty on the zenith and azimuth angles drawn from the covariance matrix.

\begin{figure}[!h]
	\centering
	\includegraphics[width=.7\textwidth]{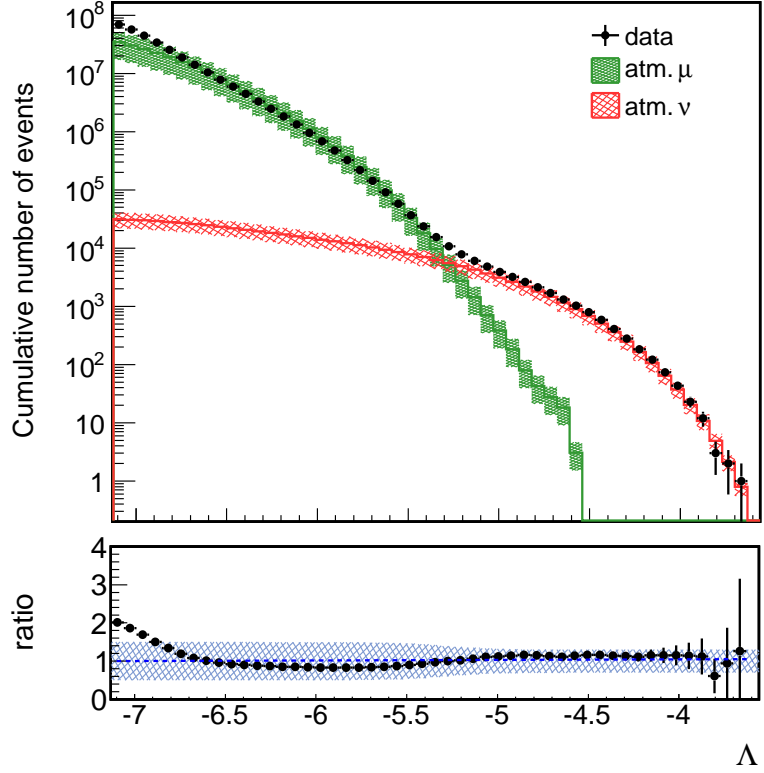}
	\caption{Cumulative distribution of the track reconstruction quality parameter, $\Lambda$, for tracks with $\cos\theta <0.1$ which have an angular error estimate $\beta< 1^\circ$. The bottom panel shows the ratio between data and simulation. The green (red) distribution corresponds to the simulated atmospheric muons (neutrinos), where a 50\% (30\%) relative error was assigned \cite[]{50per,30per}. Data errors correspond to statistical errors only.}
	\label{lambdafig}
\end{figure}

Neutrinos and atmospheric muons are simulated with the GENHEN \cite[]{GENHEN} and MUPAGE \cite[]{MUP1, MUP2} packages, respectively. Furthermore, the propagation of the muon tracks is simulated with the KM3 package \cite[]{GENHEN}. A data versus simulation comparison of the $\Lambda$ distribution for zenith angles $\theta$ with $\cos\theta < $ 0.1 can be seen in Figure \ref{lambdafig}, where the atmospheric neutrino simulation uses the Bartol flux \cite[]{BARTOL}. 

Events are selected following a blind procedure on pseudo-experiments before performing the analysis on data. The cuts on reconstructed tracks ($\Lambda > -$5.2, $\beta <$ 1$^\circ$ and $\cos\theta <$ 0.1) are chosen so that the neutrino flux needed to make a 5$\sigma$ discovery in 50\% of the experiments is minimised. This selection leads to a final data sample of 5516 events, which includes an estimated 10\% background from mis-reconstructed atmospheric muons.

\section{Detector performance}\label{performance}

For a neutrino energy spectrum proportional to E$^{-2}$, the angular resolution and acceptance for events passing the selection cuts are computed.

 \begin{figure}[!t]
	\centering
	\plottwo{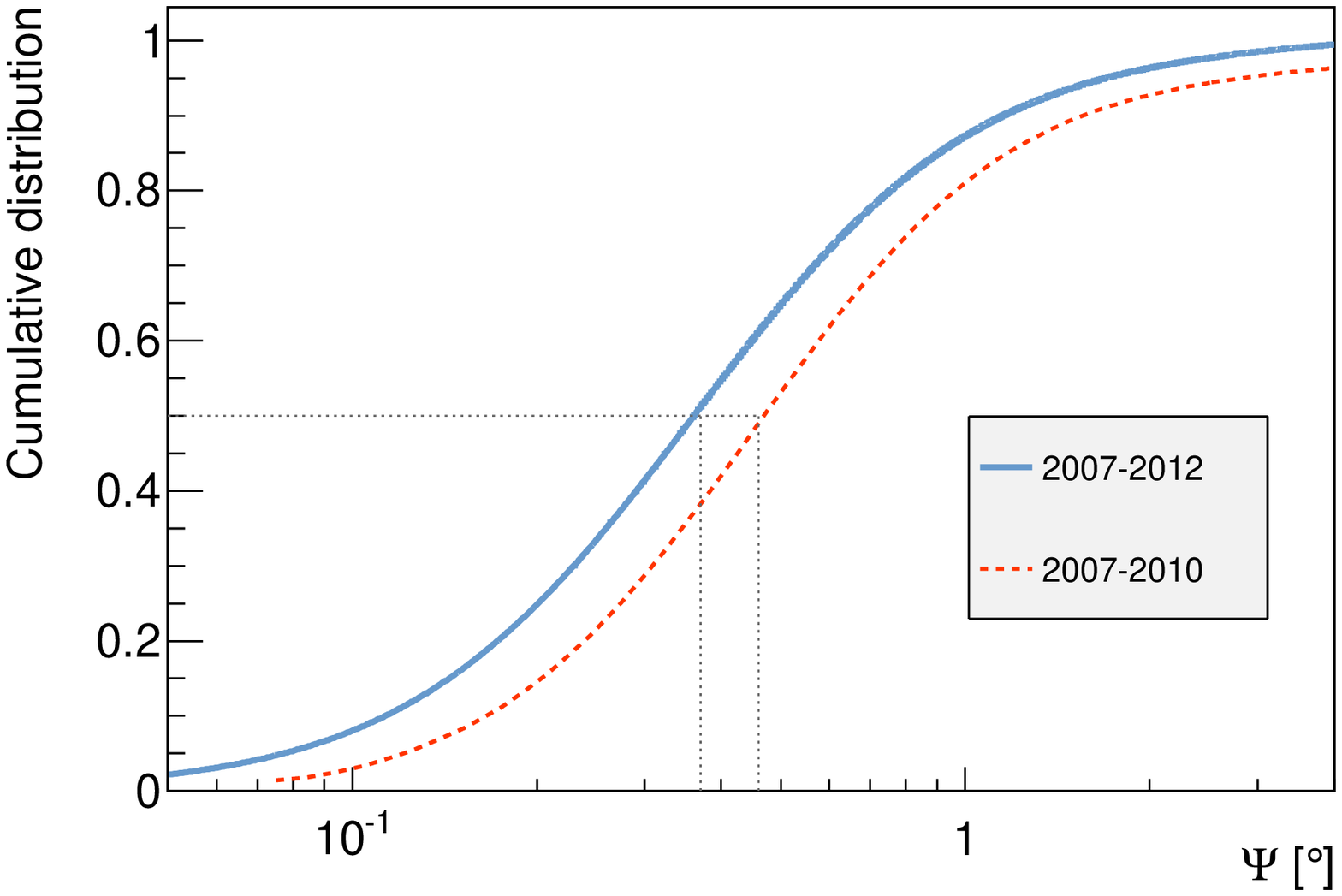}{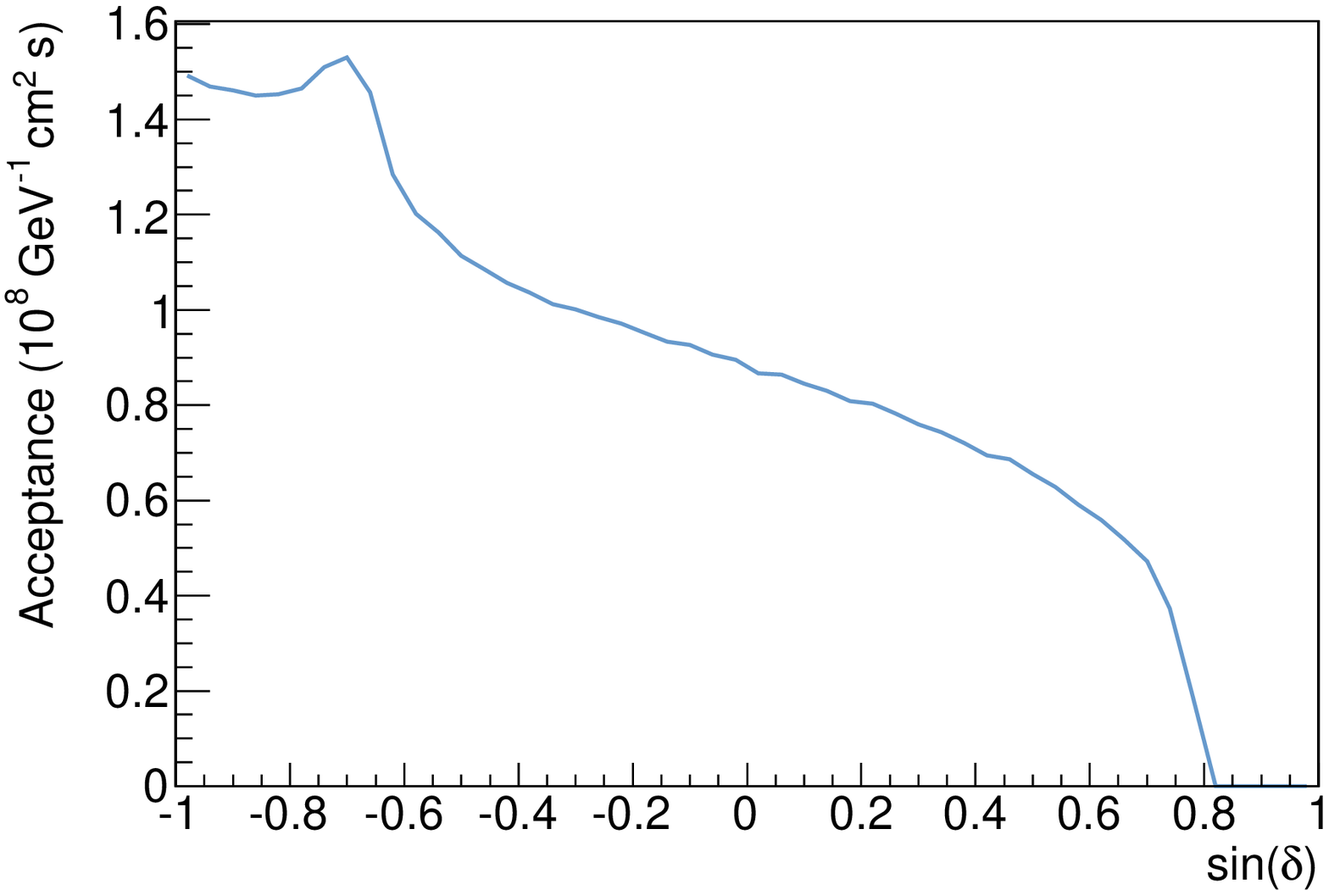}
	\caption{Left: Neutrino angular resolution determined as the median of the cumulative distribution of the reconstruction angle, $\Psi$, for the present data (solid blue line) compared to the 2007-2010 analysis (dashed red line). The black-dotted line indicates the median value. Right: Acceptance (defined in Equation \ref{eq:acc}) as a function of the declination $\delta$. An E$^{-2}$ source spectrum has been assumed for both figures.}
	\label{angres-acc}
\end{figure}
An improved modelling of the PMT transit-time distribution compared to Ref. \cite[]{PrevAna} has been used for the simulation. As a result, the estimated median neutrino angular resolution is 0.38$^\circ$, which corresponds to a 15\% improvement. Figure \ref{angres-acc} (left) shows the cumulative distribution of the angle $\Psi$ between the reconstructed muon direction and the true neutrino direction. The distribution is represented both for the whole data set (blue line) and for the previous analysis (dashed red line). 

The ``acceptance'' is defined as the quantity that multiplied by a given flux, $\Phi_0 = E^{2}_{\nu}\frac{d\Phi}{dE_{\nu}}$, gives the number of signal events. This quantity is proportional to the detector response and depends on the source energy spectrum and declination. The acceptance for a source located at a declination $\delta$ is

\begin{equation}\label{eq:acc}
A(\delta) = \Phi_0^{-1} \int dt \int dE_\nu A_{eff}(E_\nu,\delta) \frac{d\Phi}{dE_\nu} \; ,
\end{equation}

\noindent where the time integration extends over the whole period of 1338 days and $A_{eff}$ is the neutrino effective area. 
The acceptance as a function of the declination $\delta$ is shown in Figure \ref{angres-acc} (right).

\section{Search method}\label{method}

Signal events are expected to accumulate in clusters over a background of diffusely distributed atmospheric neutrinos. The search for clusters is performed using a maximum-likelihood estimation, which describes the data as a mixture of a signal and background probability density functions (PDFs):

\begin{equation}
	\log L_{s+b} = \sum_i \frac{n_s}{N}S_i + \left(1-\frac{n_s}{N}\right)B_i  .
\end{equation}

Both the background and the signal PDFs, $B_i$ and $S_i$ respectively, depend on the reconstructed direction, $\vec{x}_i$ = ($\alpha_i$, $\delta_i$) (where $\alpha_i$ and $\delta_i$ indicate the reconstructed right ascension and declination, respectively), for the $i$-th event. The parameter $n_s$ represents the expected number of signal events for a particular source and N, the total number of events in the sample. The signal PDF is defined as

\begin{equation}
S_i = \frac{1}{2\pi\beta_i^2} e^{-\frac{|\vec{x}_i-\vec{x}_s|^2}{2\beta_i^2}} P_{s}(\mathcal{N}^{hits}_i, \beta_i) ,
\end{equation}

\noindent where $\vec{x}_s$ = ($\alpha_s$, $\delta_s$) indicates the position of the source and $P_{s}(\mathcal{N}^{hits}_i, \beta_i)$ is the probability for a signal event $i$ at a position $\vec{x}_i$ to be reconstructed with an angular error estimate of $\beta_i$ and a number of hits $\mathcal{N}^{hits}_i$. The number of hits $\mathcal{N}^{hits}_i$ is a proxy for the energy of the event. 

The background PDF is described as 

\begin{equation}
B_i = \frac{B(\delta_i)}{2\pi} P_{b}(\mathcal{N}^{hits}_i, \beta_i) ,
\end{equation}

\noindent where $B(\delta_i)$ is the probability to find an event at a declination $\delta_i$ and $P_{b}(\mathcal{N}^{hits}_i, \beta_i)$ is the probability for a background event to be reconstructed with a number of hits $\mathcal{N}^{hits}_i$ and an angular error estimate of $\beta_i$.

The significance of any observation is determined by the test statistic, TS, which is defined as TS = $\log L_{s+b} - \log L_{b}$, where $L_{b}$ indicates the likelihood value for the background only case ($n_s$ = 0). Larger TS values indicate a lower probability (p-value) of the observation to be produced by the expected background. 

\section{Full sky and candidate list searches}\label{results}

A full-sky search and a search on an a pre-selected list of candidate sources are performed.

\begin{figure}[!p]
	\centering
	\includegraphics[width=.75\textwidth]{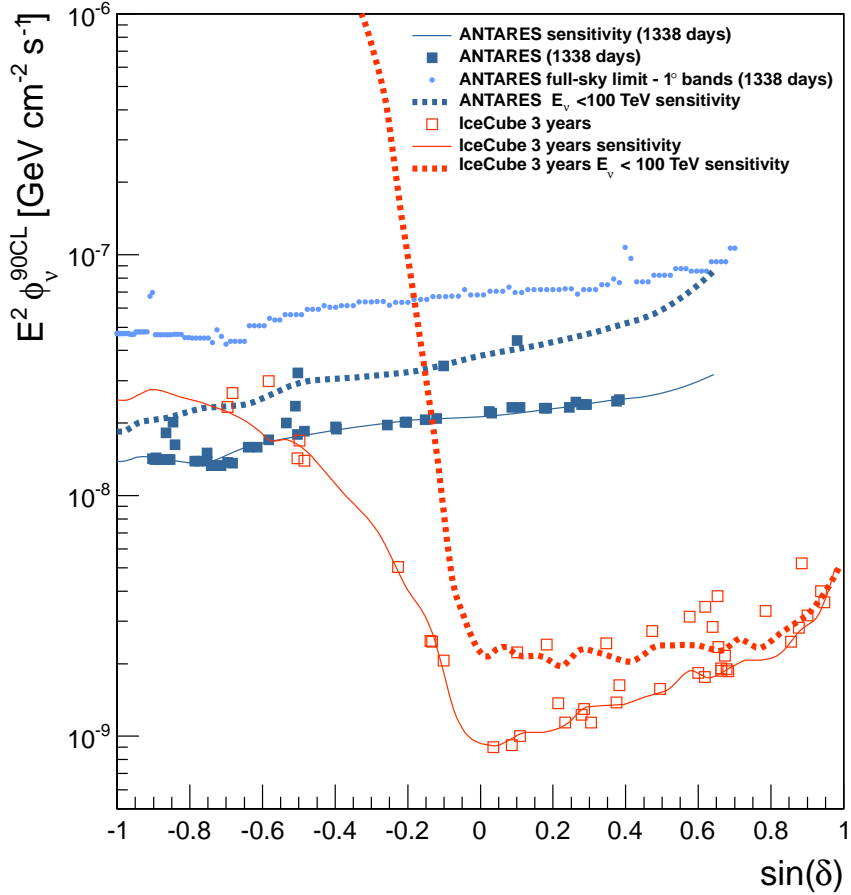}
	\caption{90 \% C.L. flux upper limits and sensitivities on the muon neutrino flux for six years of ANTARES data. IceCube results are also shown for comparison. The light-blue markers show the upper limit for any point source located in the ANTARES visible sky in declination bands of 1$^\circ$. The solid blue (red) line indicates the ANTARES (IceCube) sensitivity for a point-source with an $E^{-2}$ spectrum as a function of the declination. The blue (red) squares represent the upper limits for the ANTARES (IceCube) candidate sources. Finally, the dashed dark blue (red) line indicates the ANTARES (IceCube) sensitivity for a point-source and for neutrino energies lower than 100 TeV, which shows that the IceCube sensitivity for sources in the Southern hemisphere is mostly due to events of higher energy. The IceCube results were derived from \cite{IcePoint}. }
	\label{sensitivity}
\end{figure}

The full-sky search looks for an excess of signal events located anywhere in the whole ANTARES visible sky. A pre-clustering algorithm to select candidate clusters of at least 4 events in a cone of half-opening angle of 3$^\circ$ is performed. For each cluster, $L_{s+b}$ is maximised by variying the free parameters $\vec{x}_s$ and n$_s$. In this analysis, the most significant cluster is found at ($\alpha$, $\delta$) = ($-$46.8$^\circ$, $-$64.9$^\circ$) with a post-trial p-value of 2.7\% (significance of 2.2$\sigma$ using the two-sided convention). This direction is consistent with the most significant cluster found in the previous analysis.  
The number of fitted signal events is $n_s = $ 6.2 . A total of 6 (14) events in a cone of 1$^\circ$ (3$^\circ$) around the fitted cluster centre are found. Upper limits at the 90\% confidence level (C.L.) on the muon neutrino flux from point sources located anywhere in the visible ANTARES sky are given by the light blue-dashed line in Figure \ref{sensitivity}. Each value corresponds to the highest upper-limit obtained in declination bands of 1$^\circ$. 
 
The second search uses a list of 50 neutrino candidate-source positions at which the likelihood is evaluated. The list of sources with their corresponding pre-trial p-values and flux upper limits is presented in Table \ref{tab:CL}. The largest excess corresponds to HESS J0632+057, with a post-trial p-value of 6.1\% (significance of 1.9$\sigma$ using the two-sided convention). The fitted number of source events is $n_s$ = 1.6 . The limits for these 50 selected sources and the overall fixed-source sensitivity of the telescope are reported in Figure \ref{sensitivity}. The 90\% C.L. flux upper limits and sensitivities are calculated by using the Neyman method \cite[]{NEYMAN}.

\begin{deluxetable}{lccccclccccc}
\tabletypesize{\scriptsize}
\tablewidth{0pt}
\tablecaption{Pre-trial p-values, $p$, fitted number of source events, $n_s$, and 90\% C.L. flux limits, $\Phi^{90CL}_{\nu}$, obtained for the 50 candidate sources. The fluxes are in units of 10$^{-8}$ GeV cm$^{-2}$s$^{-1}$.}
\tablehead{
\colhead{Name}           			  & \colhead{$\alpha$ ($^\circ$)}      &
\colhead{$\delta$ ($^\circ$)}          & \colhead{$n_s$ }  &
\colhead{$p$}          & \colhead{$\phi^{90CL}_{\nu}$} & 
\colhead{Name}           			  & \colhead{$\alpha$ ($^\circ$)}      &
\colhead{$\delta$ ($^\circ$)}          & \colhead{$n_s$ }  &
\colhead{$p$}          & \colhead{$\phi^{90CL}_{\nu}$} } 
\startdata
HESSJ0632+057   & 98.24         & 5.81  & 1.60  & 0.0012  & 4.40 & HESSJ1912+101   & -71.79        & 10.15         & 0.00  & 1.00  & 2.31 \\
HESSJ1741-302   & -94.75        & -30.20        & 0.99  & 0.003  & 3.23 & PKS0426-380     & 67.17         & -37.93        & 0.00  & 1.00  & 1.59 \\
3C279   & -165.95       & -5.79         & 1.11  & 0.01  & 3.45 & W28     & -89.57        & -23.34        & 0.00  & 1.00  & 1.89 \\
HESSJ1023-575   & 155.83        & -57.76        & 1.98  & 0.03  & 2.01 & MSH15-52        & -131.47       & -59.16        & 0.00  & 1.00  & 1.41 \\
ESO139-G12      & -95.59        & -59.94        & 0.79  & 0.06  & 1.82 & RGBJ0152+017    & 28.17         & 1.79  & 0.00  & 1.00  & 2.19 \\
CirX-1          & -129.83       & -57.17        & 0.96  & 0.11  & 1.62 & W51C    & -69.25        & 14.19         & 0.00  & 1.00  & 2.32 \\
PKS0548-322     & 87.67         & -32.27        & 0.68  & 0.10  & 2.00 & PKS1502+106     & -133.90       & 10.52         & 0.00  & 1.00  & 2.31 \\
GX339-4         & -104.30       & -48.79        & 0.50  & 0.14  & 1.50 & HESSJ1632-478   & -111.96       & -47.82        & 0.00  & 1.00  & 1.33 \\
VERJ0648+152    & 102.20        & 15.27         & 0.59  & 0.11  & 2.45 & HESSJ1356-645   & -151.00       & -64.50        & 0.00  & 1.00  & 1.42 \\
PKS0537-441     & 84.71         & -44.08        & 0.24  & 0.16  & 1.37 & 1ES1101-232     & 165.91        & -23.49        & 0.00  & 1.00  & 1.92 \\
MGROJ1908+06    & -73.01        & 6.27  & 0.21  & 0.14  & 2.32 & HESSJ1507-622   & -133.28       & -62.34        & 0.00  & 1.00  & 1.41 \\
Crab    & 83.63         & 22.01         & 0.00  & 1.00  & 2.46 & RXJ0852.0-4622          & 133.00        & -46.37        & 0.00  & 1.00  & 1.33 \\
HESSJ1614-518   & -116.42       & -51.82        & 0.00  & 1.00  & 1.39 & RCW86   & -139.32       & -62.48        & 0.00  & 1.00  & 1.41 \\
HESSJ1837-069   & -80.59        & -6.95         & 0.00  & 1.00  & 2.09 & RXJ1713.7-3946          & -101.75       & -39.75        & 0.00  & 1.00  & 1.59 \\
PKS0235+164     & 39.66         & 16.61         & 0.00  & 1.00  & 2.39 & SS433   & -72.04        & 4.98  & 0.00  & 1.00  & 2.32 \\
Geminga         & 98.31         & 17.01         & 0.00  & 1.00  & 2.39 & 1ES0347-121     & 57.35         & -11.99        & 0.00  & 1.00  & 2.01 \\
PKS0727-11      & 112.58        & -11.70        & 0.00  & 1.00  & 2.01 & VelaX   & 128.75        & -45.60        & 0.00  & 1.00  & 1.33 \\
PKS2005-489     & -57.63        & -48.82        & 0.00  & 1.00  & 1.39 & HESSJ1303-631   & -164.23       & -63.20        & 0.00  & 1.00  & 1.43 \\
PSRB1259-63     & -164.30       & -63.83        & 0.00  & 1.00  & 1.41 & LS5039          & -83.44        & -14.83        & 0.00  & 1.00  & 1.96 \\
HESSJ1503-582   & -133.54       & -58.74        & 0.00  & 1.00  & 1.41 & PKS2155-304     & -30.28        & -30.22        & 0.00  & 1.00  & 1.79 \\
PKS0454-234     & 74.27         & -23.43        & 0.00  & 1.00  & 1.92 & Galactic Centre          & -93.58        & -29.01        & 0.00  & 1.00  & 1.85 \\
PKS1454-354     & -135.64       & -35.67        & 0.00  & 1.00  & 1.70 & CentaurusA      & -158.64       & -43.02        & 0.00  & 1.00  & 1.36 \\
HESSJ1834-087   & -81.31        & -8.76         & 0.00  & 1.00  & 2.06 & W44     & -75.96        & 1.38  & 0.00  & 1.00  & 2.23 \\
HESSJ1616-508   & -116.03       & -50.97        & 0.00  & 1.00  & 1.39 & IC443   & 94.21         & 22.51         & 0.00  & 1.00  & 2.50 \\
H2356-309       & -0.22         & -30.63        & 0.00  & 1.00  & 2.35 & 3C454.3         & -16.50        & 16.15         & 0.00  & 1.00  & 2.39 \\
\enddata
\label{tab:CL}
\end{deluxetable}

\section{Implications for the interpretation of the recent IceCube results}\label{Implications}

Following the recent evidence of high energy neutrinos by IceCube  \cite[]{IceCube1}, a point source close to the Galactic Centre has been proposed to explain the accumulation of seven events in its neighbourhood \cite[]{MCGonz}.
The corresponding flux normalisation of this hypothetical source ($\alpha$ = $-$79$^\circ$,  $\delta$ =  $-$23$^\circ$) is expected to be $\Phi_0 = 6 \times 10^{-8}$  \textrm{GeVcm$^{-2}$s$^{-1}$}.

This hypothetical source might be located at a different point in the sky due to the large uncertainty of the direction estimates of these IceCube events. The full sky algorithm with the likelihood presented in Ref. \cite[]{PrevAna} is used, restricted to region of 20$^\circ$ around the proposed location. The trial factor of this analysis is smaller than in the full sky search because of the smaller size of the region. In addition to the point source hypothesis, three Gaussian-like source extensions are assumed (0.5$^\circ$, 1$^\circ$ and 3$^\circ$). As in the full sky search, a half opening angle of 3$^\circ$ is used for the pre-clustering selection for source widths smaller than 3$^\circ$. In the case of the 3$^\circ$ source assumption, the angle is of 6$^\circ$.

\begin{figure}[!h]
\centering
	\includegraphics[width=.75\textwidth]{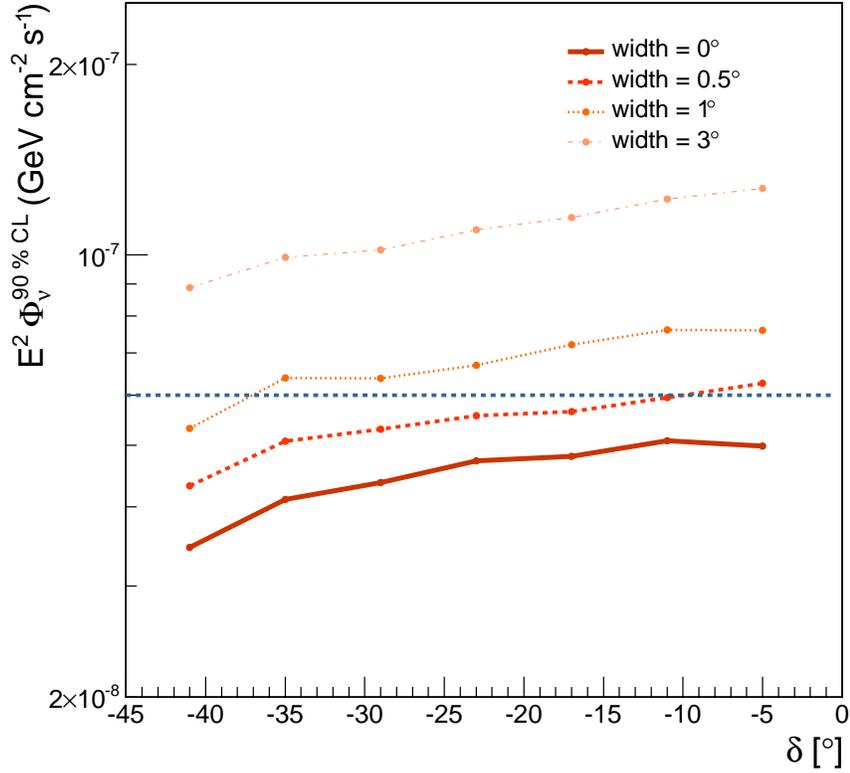}
	\caption{90 \% C.L. upper limits obtained for different source widths as a function of the declination. The blue horizontal dashed line corresponds to the signal flux given by \cite[]{MCGonz}. }
	\label{fig:Ecut}
\end{figure}

No significant cluster has been found. 
Figure \ref{fig:Ecut} shows the 90\% C.L. flux upper limits obtained for the four assumed different spatial extensions of the neutrino source as a function of the declination. The presence of a point source with a flux normalisation of \textrm{6$\times$10$^{-8}$ GeVcm$^{-2}$s$^{-1}$} anywhere in the region is excluded. Therefore, the excess found by IceCube in this region cannot be caused by a single point source. Furthermore, a source width of 0.5$^{\circ}$ for declinations lower than $-$11$^{\circ}$ is also excluded.
For an E$^{-2}$ spectrum, neutrinos with E$\;>$ 2 PeV contribute only 7\% to the event rate, hence these results are hardly affected by a cutoff at energies on the order of PeV.

\section{Conclusion}\label{conclusions}

In this paper the results of a search for neutrino point sources with six years of ANTARES data (2007-2012) are presented using two complementary analyses. Firstly, a scan for point sources of the ANTARES visible sky. Secondly, a search for correlations of events with a pre-selected  list of candidate sources for neutrino emission. In the first case, the most significant cluster has a post-trial p-value of 2.7\% (a significance of 2.2$\sigma$). In the case of the candidate list study, the largest excess corresponds to HESS J0632+057 with a post-trial p-value of 6.1\% (1.9$\sigma$). Both results are compatible with a pure background hypothesis. The derived flux upper limits are the most restrictive in a significant part of the Southern sky. The possibility that the accumulation of 7 events reported by IceCube near the Galactic Centre is produced by a single point source has been excluded. These results show the potential of neutrino telescopes in the Northern hemisphere, such as the planned KM3NeT observatory \citep[]{KM3Net}, to interpret the increasing evidence of cosmic neutrino fluxes.

\acknowledgments

The authors acknowledge the financial support of the funding agencies:
Centre National de la Recherche Scientifique (CNRS), Commissariat \`a
l'\'Ene\-gie Atomique et aux \'Energies Alternatives (CEA), 
Commission Europ\'eenne (FEDER fund
and Marie Curie Program), R\'egion Alsace (contrat CPER), R\'egion
Provence-Alpes-C\^ote d'Azur, D\'e\-par\-tement du Var and Ville de La
Seyne-sur-Mer, France; Bundesministerium f\"ur Bildung und Forschung
(BMBF), Germany; Istituto Nazionale di Fisica Nucleare (INFN), Italy;
Stichting voor Fundamenteel Onderzoek der Materie (FOM), Nederlandse
organisatie voor Wetenschappelijk Onderzoek (NWO), the Netherlands;
Council of the President of the Russian Federation for young
scientists and leading scientific schools supporting grants, Russia;
National Authority for Scientific Research (ANCS), Romania; Ministerio
de Ciencia e Innovaci\'on (MICINN), Prometeo of Generalitat Valenciana
and MultiDark, Spain; Agence de l'Oriental and CNRST, Morocco. We also
acknowledge the technical support of Ifremer, AIM and Foselev Marine
for the sea operation and the CC-IN2P3 for the computing facilities.

{\it Facilities:} \facility{ANTARES}.

\end{document}